\documentclass[prl,twocolumn]{revtex4}
\usepackage{amsfonts}
\usepackage{amsmath}
\usepackage{psfrag}
\usepackage{amssymb,amsmath,amsthm}
\usepackage[dvips]{graphicx}
\usepackage{verbatim} %
\usepackage{color}
\newcommand{\beq}{\begin{equation}}
\newcommand{\eeq}{\end{equation}}
\newcommand{\bea}{\begin{eqnarray}}
\newcommand{\eea}{\end{eqnarray}}
\usepackage{color}

\newcommand{\bma}{\left(\begin{matrix}}
\newcommand{\ema}{\end{matrix}\right)}
\def\build#1_#2{\mathrel{\mathop{#1}\limits_{#2}}}
\definecolor{pink}{rgb}{1,0.5,0.5}
\definecolor{violet}{rgb}{1,0,1} 
\definecolor{red}{rgb}{1,0,0}
\definecolor{yellow}{rgb}{0.7,1,0}
\definecolor{orange}{rgb}{1,0.5,0}
\definecolor{white}{rgb}{1,1,1}
\definecolor{blue}{rgb}{0,0,1}
\definecolor{cyan}{rgb}{0,1,1}

\begin{document}

\begin{abstract}
The energy of a dislocation loop in a continuum elastic solid under pressure is considered within the framework of classical mechanics. For a circular loop, this is a function with a maximum at pressures that are well within reach of experimental conditions for solid helium suggesting, in this case, that dislocation loops can be generated by a pressure-assisted thermally activated process. It is also pointed out that pinned dislocations segments can alter the shear response of solid helium, by an amount consistent with current measurements, without any unpinning.
\end{abstract}

\title{Shear Modulus of an Elastic Solid under External Pressure \\ as a function of Temperature: The case of Helium}

\author{Felipe Barra$^1$, Fernando Lund$^1$, Nicol\'as Mujica$^1$ and Sergio Rica$^2$}

\affiliation{\mbox{$^1$Departamento de F\'\i sica and CIMAT, Facultad de Ciencias
F\'\i sicas y Matem\'aticas, Universidad de Chile, Santiago, Chile} \\
$^2$Facultad de Ingenier\'\i a y Ciencias, Universidad Adolfo Ib\'a\~nez, Santiago, Chile.}

\maketitle

\paragraph{Introduction.} Dislocations in a solid, when present in sufficient number, change its elastic properties \cite{Dislocations}. This classic subject has been recently revisited \cite{M4,Maurel2009D} with the aim of using acoustics as a nonintrusive probe of plasticity in metals \cite{Barra2009}. In the early days of dislocation theory there was an interest in the pressure-assisted thermal generation of dislocation loops \cite{friedel}. Interest in this topic decayed because the needed pressures would be much too big for usual engineering materials. Solid helium, however, offers the possibility of laboratory measurements to test these ideas, since in this case experimental pressures are a significant fraction of the shear modulus and, as we point out below, have measurable effects in the experimental temperature range.  Indeed, recent experimental results on the mechanical properties of  solid helium have brought the influence of dislocations to the fore. 

At temperatures Êon the order of 100 mK, torsional oscillator experiments suggest that solid helium may display  superfluid properties \cite{chan,rittner,Kojima,Kubota,Kondo,kim,Reviews}, theoretically predicted more than forty years ago \cite{andreev, reatto,leggett}. Mechanical properties of solid helium in that regime have also been probed: Ê``almost static'' strain-stress \cite{Day2007}, as well as acoustic Ê\cite{beamprl2010,balibar} and dc rotation \cite{kim} measurements have shown that the shear modulus of Êsolid helium presents an anomaly at low temperatures. Indeed, starting at about 70 mK, as the temperature increases the shear modulus abruptly decreases to a constant value that is 4\% to 10\% smaller, Êat a temperature on the order of 200 mK. It has been suggested \cite{Day2007,ChanImp2008,beamprl2010} that the presence of dislocations may be responsible of this behavior: at low temperatures $^3$He impurities are strongly pinned and they, in turn, pin the dislocations that are regarded as a stiff network. As temperature increases the $^3$He impurities are released, thus unpinning the dislocations that become mobile, and the solid softens. While this is a physically appealing picture, in this Letter we point out additional properties of dislocations, within a classical mechanics framework, that provide further insight into the observed shear modulus anomalies. One of them stems from the fact that
solid helium exists only under pressure, at a level that is at least on the order of 15\% of its shear modulus, and can go considerably over that value. The other is that, even when strongly pinned, a dislocation network will alter the shear response of a solid.

An important aspect of the debate concerning the role of dislocations is to understand to what extent some of the available data can be understood on purely classical grounds. Although helium is a quantum solid, from a mechanical point of view the quantum character can be manifested computing a dimensionless analog of de Boer parameter, in which the interaction energy is estimated through the shear modulus:
$$ 
\Lambda_\mu = \frac{\hbar^2}{m_{\rm He} a^5 \mu},
$$
where $\hbar $ is Planck's constant, $a$ is a typical microscopic Êlength, say the lattice constant,
$m_{\rm He}$ is the atomic mass, and $\mu$ the shear modulus at standard conditions \cite{data}.
This gives $ \Lambda_\mu \approx1.6 \times Ê10^{-2} $ for $^4$He, to be compared with ÊÊ$3.8 \times 10^{-3}$ for solid hydrogen, Ê$3 \times 10^{-5} $ for neon, and $3.5 \times 10^{-7} $ for copper. The smallness of $\Lambda_\mu$ indicates that a classical treatment of the mechanical properties of solid helium is a reasonable first approximation.

\begin{figure}[h!]
\includegraphics[width=.8\columnwidth]{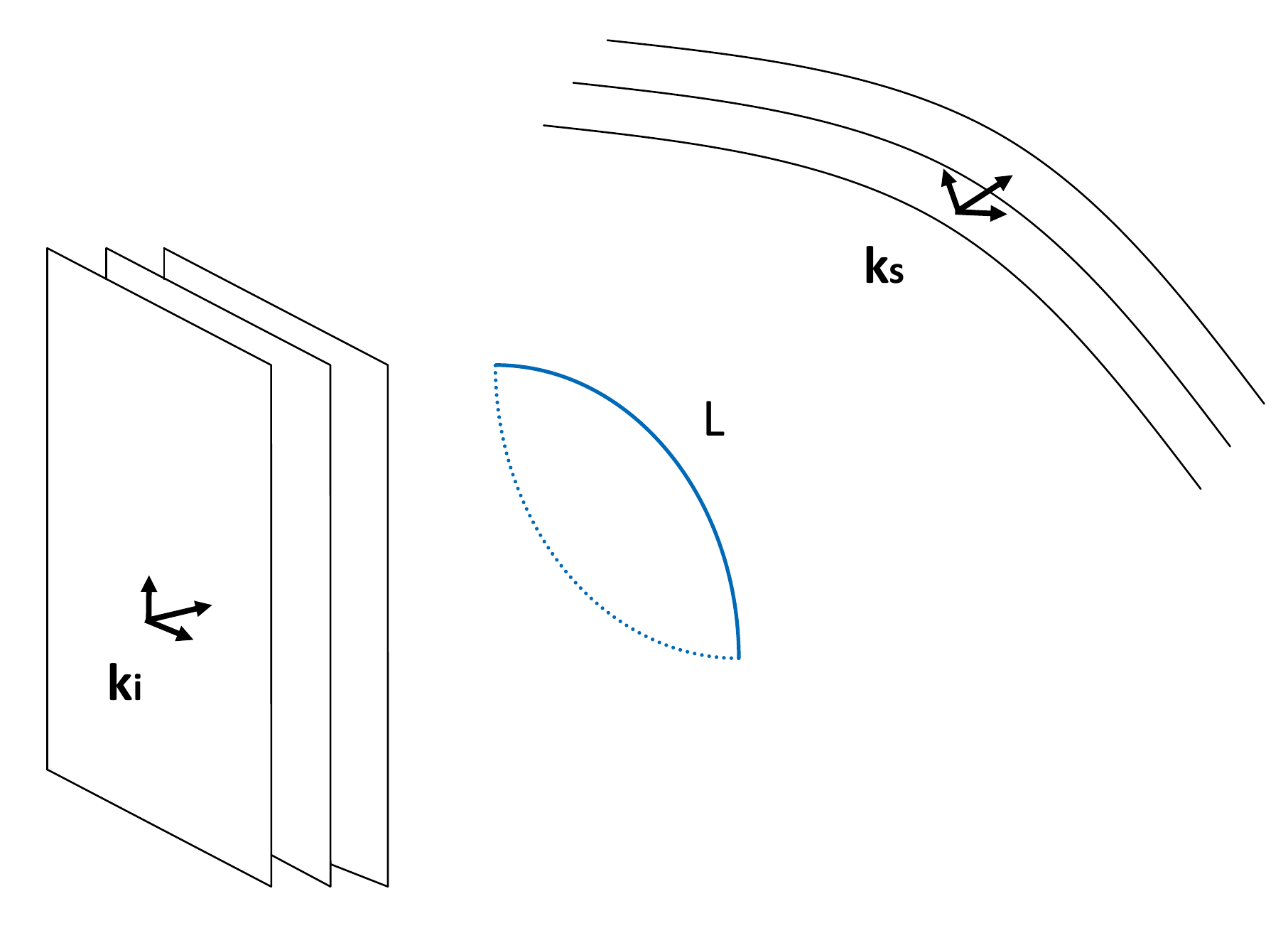}
\caption{The basic scattering mechanism of an elastic wave by an oscillating line dislocation: an  elastic wave of wave vector $\vec k_i$ is incident upon a dislocation segment of length $L$ that oscillates in response. As it does, it re-emits waves with scattered wave vector $\vec k_s$.}
\label{fig_scatt1}
\end{figure}

\paragraph{Scattering of elastic waves by dislocation segments.}  Maurel et al. \cite{M4}, using multiple scattering theory, have generalized the Granato-L\"ucke theory \cite{granatolucke} to study the coherent propagation of waves in an elastic medium that is filled with dislocation segments, pinned at their ends, and with random locations, orientations, equilibrium lengths, and Burgers vectors. These (vector) dislocation segments oscillate like strings when forced by an elastic wave (Figure \ref{fig_scatt1}).

Consider a dislocation segment $\vec X (s,t)$ of length $L$ at equilibrium,  moving at
low velocities (i.e., small compared with the speeds of elastic waves) with pinned ends, in an isotropic, homogeneous, elastic continuum of density $\rho$, (bare) Lam\'e parameters $\lambda$ and $\mu$ and corresponding longitudinal and transverse wave velocities $c_{L} = \sqrt{(\lambda + 2 \mu)/\rho}$ and $c_{T} = \sqrt{\mu /\rho}$. Low accelerations are also assumed, so that the back-reaction of the radiation on the dislocation dynamics can be
neglected. Following Ref. \cite{lund88}  and under these hypothesis, the
equation of motion of an
edge dislocation takes the form of the equation of motion for a string
endowed with mass and line tension, forced by the usual Peach-Koehler
force~\cite{PeachKoehler,koehler}
\begin{equation}
m \ddot{X}_k(s ,t) - B\dot{X}_k(s ,t) - \Gamma X_k''(s ,t)= F_k,
\label{eqMouv}
\end{equation}
and the associated boundary conditions at pinned ends $X_k(\pm L/2,t)=0$ (see Fig. \ref{fig_scatt1}). In
Eqn. (\ref{eqMouv}) 
\begin{equation}
m \equiv \frac{\rho b^2}{4\pi}(1+\gamma^{-4}) \ln(\Lambda/\Lambda_0)  \sim \rho b^2,
\label{defmasse}
\end{equation}
defines a mass per unit length (with $\Lambda$, $\Lambda_0$
the long- and short-distance cut-off lengths, respectively) and
\begin{equation}
\Gamma \equiv
\frac{\rho b^2}{2\pi} (1-\gamma^{-2})  c_{T}^2 \ln(\Lambda/\Lambda_0) \sim \mu b^2
\label{deflinetension}
\end{equation}
is the line tension, $\gamma \equiv c_{L}/c_{T} = [2(1-\nu)/(1-2\nu)]^{1/2}$, with $\nu$ the Poisson's ratio, and $b =|\vec b|$ where $\vec b$ is the Burgers vector. $B$ is the viscous drag coefficient, and 
  $F_k = \epsilon_{kjm}
  t_m b_i  \sigma_{ij}$  is the Peach-Koehler force with $t_m$ the unit tangent along the dislocation segment ($\epsilon_{ijk}$
  denotes the usual completely antisymmetric tensor). A similar expression is valid for screw dislocations. In the following, the mass term will be ignored, since only overdamped dislocation motion will be considered, an approximation valid for frequencies $\omega$ small compared to the lowest resonant frequency of the dislocation segment, $\omega \ll \omega_1 \equiv  ( \Gamma/m)^{1/2}\pi/L$, where
$\omega_1/(2\pi)$ is of the order of GHz for helium. Equivalently, this is valid for wavelengths $\lambda$ very large compared to the dislocation distance between pinning points: $\lambda \gg L$.

\paragraph{Effective elastic constants.} When many of these dislocations are present, with probability $p(L) dL$ of having a length between $L$ and $L + dL$, Maurel et al. \cite{M4} used multiple scattering theory to compute an effective, complex index of refraction whose real part gives a renormalized velocity of propagation for both longitudinal and transverse waves. In the latter case the result is
\beq
c^{\rm eff}_T  =  c_{T} \int
\left[ 1 -
  \frac{ \delta}{1+(\omega t_B )^2}   \right] p(L) dL,
\label{vitesseT2wd}
\eeq
from which an effective shear modulus follows:
\beq
\mu^{\rm eff}  =   \mu \int
\left[ 1 -
  \frac{2 \delta}{1+(\omega t_B )^2}   \right] p(L) dL.
\label{vitesse2wd}
\eeq
The imaginary part gives attenuation (in units of inverse distance) from which a quality factor can be extracted:
\beq
Q_T^{-1}   =  \int
  \frac{2 \delta}{1+(\omega t_B )^2} \left[
  (\omega t_B) +   {\mathcal O} \left( \frac{L}{\lambda}  \right)^3  \right]  p(L) dL
\label{Lambda2d}
\eeq
where 
\beq
\delta \equiv (4/5\pi^4)
  (\mu b^2 /\Gamma) n L^3 \sim  n L^3 ,\label{delta}
\eeq 
$n$ is the number of dislocations of length $L$ per unit volume, $\lambda$ is wavelength, and $t_B \equiv (BL^2 /\pi^2 \Gamma)$ is a relaxation time related to the parameters determining the dislocation dynamics. Similar expressions can be obtained for longitudinal waves. Expressions such as (\ref{vitesse2wd}) and (\ref{Lambda2d}) have been considered by Beamish et al. \cite{beamprl2010}, with relaxation times corresponding to thermally activated relaxation processes with a log-normal distribution of excitation energies. 

The reasoning above suggests that the shear modulus behavior studied by Beamish could, at least in part, be attributed to the dynamical response of dislocation segments. But, where would the dislocations come from? They may be an artifact of preparation. On the other hand,  as we show below, they can also be generated by a thermally activated process, with an activation energy that depends on the external pressure.

\begin{figure}[t!]
\includegraphics[width=.8\columnwidth]{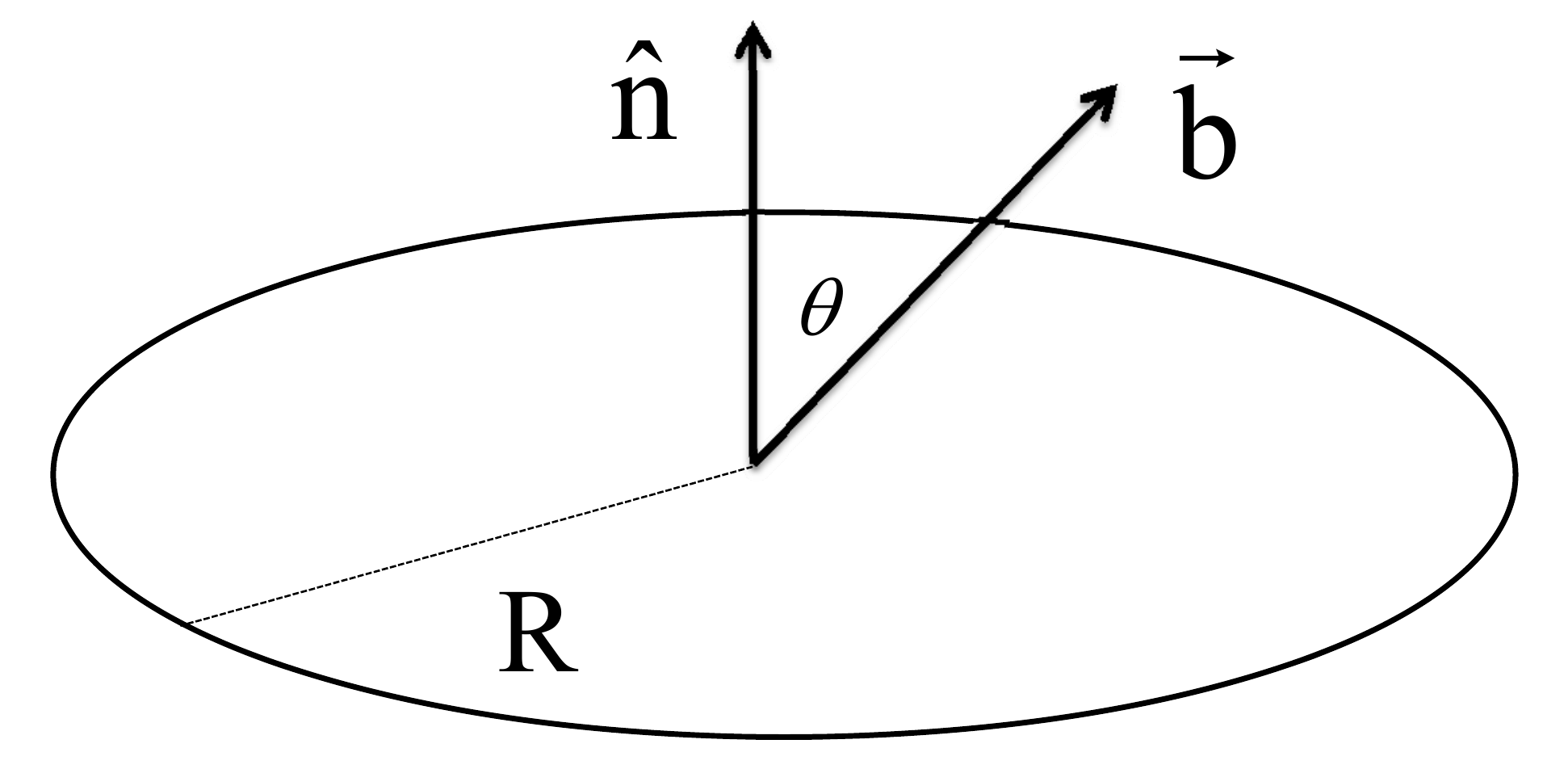}
\caption{Geometry of a dislocation loop under pressure. See Eqn. (\ref{loopenergyP}).}
\label{fig2loop}
\end{figure}

\paragraph{Thermal activation of a single dislocation loop.}  Consider a circular dislocation loop of radius $R$, core radius $\tau$ (Fig. \ref{fig2loop}) in the presence of an external pressure $\Delta P$ above the liquid-solid transition pressure $P_{LS} (\approx 25$ bar). For $R \gg \tau$ the energy of the loop is~\cite{nabarrobook}
\beq
U (R, \theta  )  = V \sin^2 \theta + W \cos^2 \theta + {\mathcal C} \cos \theta,
\label{loopenergyP}
\eeq
where $\theta$ is the angle between the Burgers vector and the normal to the plane that contains the loop,
\bea
V(R ) & = & \frac{\mu_0 (2-\nu)b^2R}{4(1-\nu)}  \left[\ln \left(\frac{R}{\tau}\right) + C_1\right ]  , \\
W(R) & = & \frac{\mu_0 b^2R}{2(1-\nu)} \left[\ln \left(\frac{R}{\tau}\right) + C_2\right ] , \\
{\mathcal C} (R)  & = & \pi R^2 b  \, \Delta P ,
\eea and $C_1$, $C_2$ are constants of order one; they provide the energy it takes to create a loop of minimal radius $R=\tau$. We shall use this formula even for small radii, respecting $R \ge \tau$. Note that we take as $\Delta P$ the excess pressure above the minimum value needed to have a solid.

 It has a global minimum on the circle $R=\tau$, and a saddle point on the axis $\theta = \pi$ ($\Delta P>0$). The location of this saddle point provides an energy barrier to be overcome by a thermal activation process. Moreover, we shall see next that the curvature at the saddle provides the pre-factor of the activation time.

The saddle is located at $R_*$, the solution of
\beq R_* =
\frac{ b \mu}{4\pi (1-\nu) \Delta P } \left(\ln\left(\frac{R_*}{\tau}\right)+ C_2 +1\right), \label{saddle}
\eeq
and the corresponding energy is
\begin{equation*}
 U(R_*,  \pi ) 
     =  \frac{\mu^2 b^3 }{16\pi (1-\nu)^2 \Delta P} \left( \left[\ln \left(\frac{R_*}{\tau}\right) + C_2\right]^2 -1  \right) .
\label{Usaddle}
\end{equation*}

Thus, when $\Delta P>0$, there is a significant change in the physics of dislocation generation, compared to the $\Delta P=0$ case: at any temperature there will now
 be a finite rate of dislocation loop generation given by an Arrhenius-like expression. A similar physics was considered by Langer and Fisher when studying the influence of an external flow in the generation of vortex loops in superfluid helium \cite{lf}.

 The barrier height $\Delta U \equiv U(R_*, \pi )  - U(\tau, \pi )$ is given by
 \beq
\Delta U = U(R_*, \pi ) - \left( \frac{\mu b^2\tau}{2(1-\nu)} C_2 - \pi \Delta P b \tau^2  \right) \label{DeltaU}
\eeq
and, at temperature $T$ there will be a generation of dislocation loops at a rate $f$ given by
\beq
f = f_0 e^{-\Delta U /k_B T},
\label{arrhenius}
\eeq
where $f_0$ is a microscopic frequency which we take to be of order of $c_T/b \approx 10^{12}$ s$^{-1}$. 

\vspace{1em}

Let us look at the conditions for the barrier to disappear, that is, when the saddle and the minimum collide, $R_* = \tau$. In this case, from (\ref{saddle}) one has that the critical pressure is 
$$  
\Delta P = \mu \frac{ C_2 +1 }{4\pi (1-\nu) } \frac{b}{\tau} \approx  0.21  \mu .
$$ 
This is an excess external pressure of order 20\% of the shear modulus. Note that, at this level, we do not consider the possible dependence of elastic constants and lattice parameters on external pressure.

\begin{figure}[t!]
a) \includegraphics[width=.9\columnwidth]{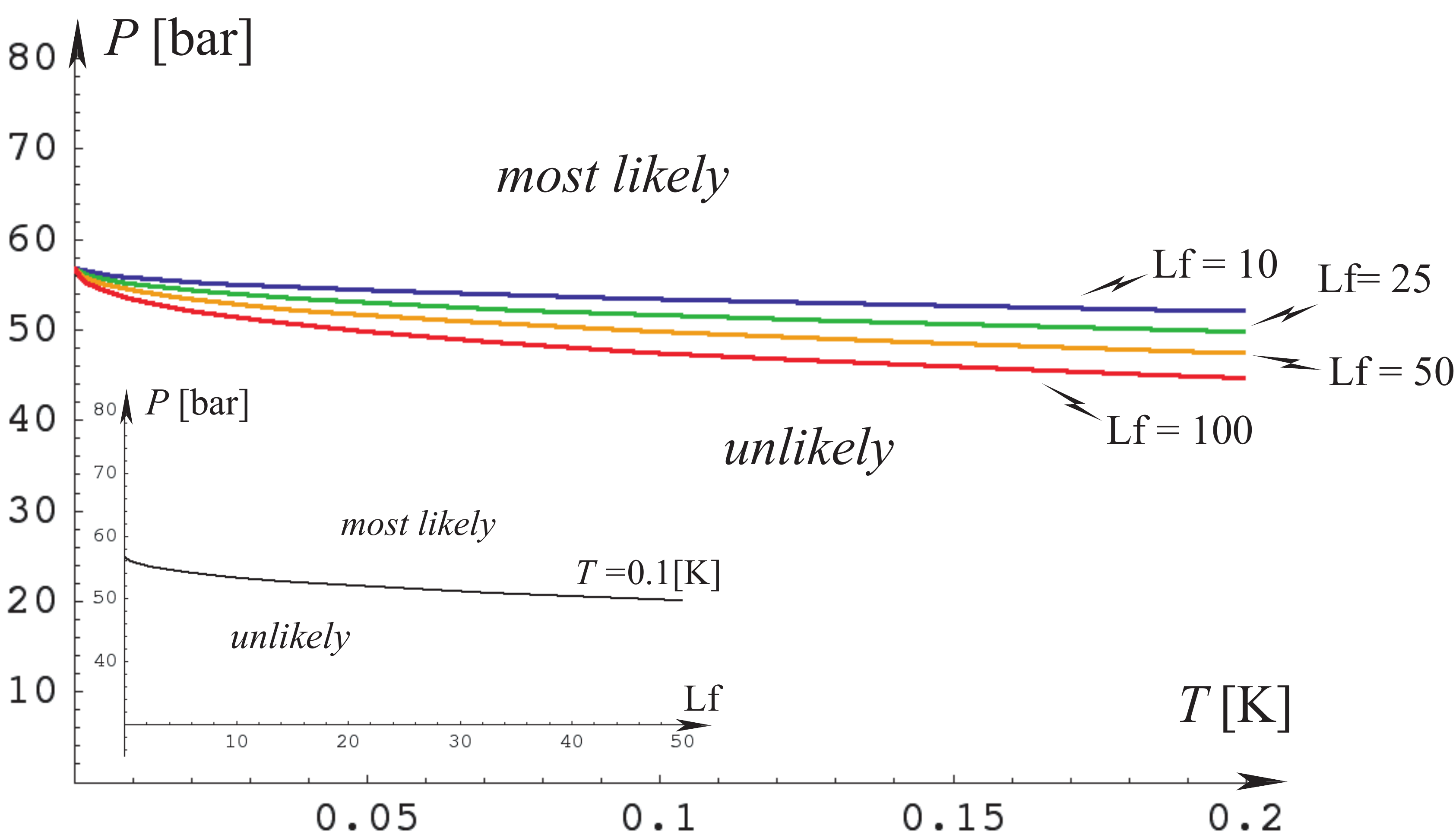}\\
b) \includegraphics[width=.9\columnwidth]{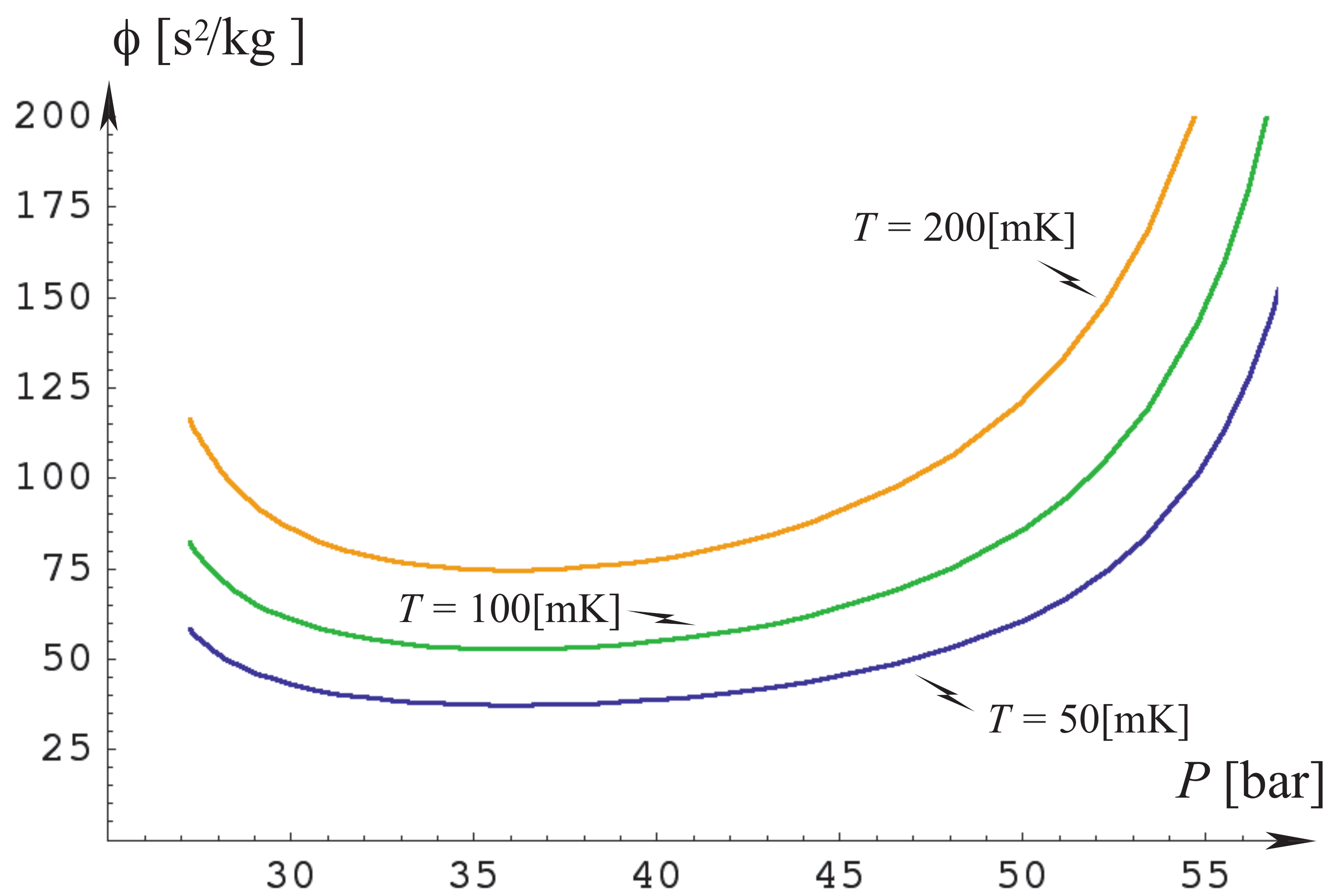}
\caption{ a) Plot of the most likely-unlikely critical pressure $P$ for dislocation loop generation vs. temperature, after (\ref{f0vsf}). The inset plots the most likely-unlikely critical pressure as a function ${\rm Lf} = \log( f_0/f )$\cite{data}. b) The value of $\Phi$, the function appearing in the pre-factor of Arrhenius' law Eq.(\ref{exittime1})  as a function of pressure $P$ for the values of temperature $T=50, \, 100, \, 200$ mK \cite{data}. Both figures plot  the actual imposed pressure $P = P_{LS} + \Delta P $ with $P_{LS}=25$ bar, for ease of visualization.}
\label{pvsT}
\end{figure}

For a barrier of finite height, the Arrhenius formula (\ref{arrhenius}) provides a relation between $\Delta P$, $T$ and the rate of dislocation loop generation:
\beq
\frac{\Delta U }{k_B T}  = \ln \left( \frac{f_0}{f} \right) \equiv {\rm Lf}.
\label{f0vsf}
\eeq 
Hence, because $\Delta U$ depends explicitly on $\Delta P$ it is possible to trace a $P-T$ curve separating the regions of large and small dislocation loop generation rate. This family of curves is plotted in Fig. \ref{pvsT}-a. It gives the nucleation pressure as a remarkably flat function of temperature, at around 50 bar.

\paragraph{Derivation of exit time.}
Next, we infer from the frequency rate for
dislocation nucleation a drag coefficient that controls
dislocation growth.
From Eqn. (\ref{loopenergyP}), dislocations loops with Burgers vector parallel to the normal to the dislocation loop ($\theta = \pi$) have an energy
$U (R) \equiv W (R) -{\cal C} (R)$. We assume that, due to thermal exitations, $R$ will satisfy a Langevin-type equation 
\begin{equation}
{\bar B} \dot{R}=-\frac{\partial U}{\partial R}+\xi(t)
\end{equation}
with $\xi(t)$ a white thermal noise $\langle \xi(t) \xi(t^\prime)\rangle=2{\bar B}k_BT\delta(t-t^\prime)$ and ${\bar B}$ a drag coefficient. Note that, in the configuration considered, the loop is a prismatic dislocation loop whose radius will increase by climb so this drag will differ qualitatively from the $B$ coefficient that appears in Eqn. (\ref{eqMouv}), that describes glide, conservative, motion. 
Also, both coefficients have different units: $B\sim$ kg/(m~Ês) and $\bar B \sim$ kg/s.

An asymptotic approximation for the exit time from the barrier can be obtained with little modification of the case
in which the potential has a minimum and a saddle \cite{vanKampen}. In the present case the potential  grows linearly at $R=\tau$ and, since $R$ can not be smaller than this value, we take  a reflecting boundary condition at $R=\tau$ for the exit time computation \cite{vanKampen}. In the small temperature limit the mean first passage time $f^{-1}$ through the saddle gives the following approximation

\begin{equation} 
\frac{ 1}{f}=\frac{\bar B }{|U'(\tau)|}\, \sqrt{\frac{2\pi\, {k_BT}}{\left| U''(R_*)\right| }  }  \times {e^{\frac{\Delta U }{k_B T }}}.
\label{exittime1}
\end{equation}
Evaluating the derivatives of the potential we get for the prefactor to the Arrhenius law Eq.(\ref{exittime1}), $f_0^{-1}  =    {\bar B} \Phi$, with
\begin{equation*}
\Phi =   \frac{1}{ b \left| \frac{(1+C_2) }{2(1-\nu)} \mu b -2\pi\tau \, \Delta P\right|} \sqrt{\frac{2\pi\, k_BT}{b \left|-2\pi\, \Delta P + \frac{\mu b}{2(1-\nu) R_*}\right|}},
\end{equation*}
 where it is understood that $R_*$ depends on $\Delta P$ through  (\ref{saddle}). The plot of $\Phi$ as a function of $\Delta P $ is given in Fig. \ref{pvsT}-b.

One may estimate $\bar B$  taking $f_0 \sim c_T/ b \approx 10^{12}$ s$^{-1}$ and $\Phi \approx 60$~s$^2$/kg, to  get 
\beq
\frac{\bar B}{\tau}\sim \frac{1}{c_T \Phi}  \approx 5 \times  10^{-5} \,\,\,\, {\rm Pa \cdot s},
\label{dragclimb}
\eeq
which coincides with typical values for dislocation drag in ordinary metals at room temperature \cite{z}.

\paragraph{Discussion.}
We have shown that the low shear modulus of solid helium makes it possible to achieve pressure-driven thermally activated generation of dislocation loops at pressures that appear to be achievable in the laboratory. Their presence, according to the multiple scattering theory of elastic waves by dislocations, can give rise to a change in shear modulus and to a quality factor of the type that has been considered by Beamish and collaborators \cite{beamprl2010} {and by recent interpretation in terms of a complex rheology \cite{hunt,pratt}}. An Arrhenius formula for the rate of dislocation generation is in agreement with available data. ÊHowever, the corresponding calculation has been carried out for an isolated dislocation loop that, in the absence of interactions with impurities, grain boundaries, or other dislocations, will grow without limit. Therefore the model, cannot, by itself, account for the saturation of the medium with dislocations. {Something,} Êat some point, must stop the dislocation generation. How will this happen?

The presence of $^3$He impurities may prevent an unlimited increase of dislocation loops  because as a dislocation touches an impurity it is pinned, thus the extension energy required becomes higher. A thermodynamical equilibrium state dominated by dislocation is then possible. Let us assume  that the mean distance between pinning points, $L$, is given by the mean distance between $^3$He impurities or, $L \sim 10^2 \tau$ to $10^3 \tau$, depending if their concentration is 1 ppM or 1 ppb, respectively. If the dislocations segments covered all bonds associated with a simple cubic lattice of $^3$He atoms, we would have $ n L^3 \sim 3$.

According to (\ref{vitesse2wd}) and (\ref{delta}), and the discussion in \cite{beamprl2010}, there is a change in shear modulus as a function of temperature given by \cite{data}
\beq
\frac{\Delta \mu}{\mu} \approx  0.02\, nL^3 
\eeq 
which, under the previous assumption gives a change of about 6\%, which is in agreement with current experimental results. Different geometrical arrangements could explain easily observed variations of this ratio.

Finally, the drag coefficient $B$  can be estimated  via  the relaxation time $t_B$, which is, after Beamish and co workers \cite{beamprl2010}, of order 9 ns. Thus  $ B = \pi^2 \Gamma t_B/ L^2 \approx   10^{-4} - 10^{-6}$ Pa s depending on the value of $L$. This is not very different from the ``climb'' drag estimated in~(\ref{dragclimb}), and consistent with the smaller value for ``glide'' drag compared to ``climb'' drag that one would expect on classical grounds. No quantum modeling of dislocation drag appears to be available.
 
To conclude, we have shown that pinned dislocation segments can significantly alter the shear response of solid helium even without their unpinning, and that external pressure may generate thermally excited dislocation loops at a significant rate under experimentally realistic conditions. This suggests that it could be interesting to perform systematic measurements of the shear response of solid helium as a function of pressure.

\paragraph{Acknowledgements.}
This work was supported by Fondecyt Grants 1110144 (FB), 1100289 (SR), and 1100198 (FL and NM). We also acknowledge Anillo grant ACT 127 (FB, FL and NM). SR is on leave of absence from Institute Nonlin\'eaire de Nice, CNRS, France.


\begin{thebibliography}{99}
\bibitem{data} The following numerical values are used throughout this paper: $a=b=\tau=0.367 \,\rm nm$, $m_{\rm He} = 6.7 \times 10^{-27} \,\rm kg$, $\mu = 150 \,\rm bar$. $\rho = 195 $ kg/m$^3$,  $\nu=1/4$, $C_2=1$ and $\Lambda/\Lambda_0= L/\tau$.
\end{thebibliography}
\end{document}